\documentclass[letter,twocolumn]{jpsj2} 

\usepackage{color,bm}

\def\runauthor{N. \textsc{Furukawa, M. \textsc{Mochizuki}}
}

\title{
Roles of Bond Alternation in Magnetic Phase Diagram of $R$MnO$_3$
}

\author{
  Nobuo \textsc{Furukawa}$^{1,2}$ and Masahito \textsc{Mochizuki}$^3$
}

\inst{
  $^{1}$Department of Physics and Mathematics, Aoyama Gakuin University,
  5-10-1 Fuchinobe, Sagamihara 229-8558, Japan\\
  $^{2}$Multiferroics Project, ERATO,
  Japan Science and Technology Agency, \\
  c/o Department of Applied Physics, The University of Tokyo, 
  7-3-1 Hongo, Tokyo 113-8656, Japan\\
  $^{3}$Department of Applied Physics, The University of Tokyo, 
  7-3-1 Hongo, Tokyo 113-8656, Japan
}

\abst{
  In order to investigate the nature of the antiferromagnetic structures 
  in perovskite $R$MnO$_3$, we study a Heisenberg
   $J_1$-$J_2$ model with bond alternation using analytical and numerical approaches.
  The magnetic phase diagram 
   which includes incommensurate spiral states and commensurate collinear states is reproduced. We discuss that
  the  magnetic structure with spin $\uparrow\uparrow\downarrow\downarrow$ configuration
 (E-type structure)  and  the ferroelectricity emerge cooperatively to stabilize this phase. Magnetoelastic couplings are crucial to
understand the magnetic and electric phase diagram of $R$MnO$_3$.
}  

\kword{
    Heisenberg model, bond alternation, E-type structure, spin spiral state,
    A-type structure, $R$MnO$_3$, multiferroics
}

\begin{document}
\maketitle

Strongly correlated electron systems show various phase transitions
involving complex order parameters, in general.
Typically, in perovskite manganites 
$R$MnO$_3$, phases with
spin, charge, orbital and lattice degrees of freedom
emerges and competes with each other,
through control of 
carrier concentrations and ionic radius of $R$ ions.\cite{Tokura00}
In these Mn ions, 3d electrons 
exhibit varieties of  phases in
nearly identical systems with a slight change in a few set of parameters.

One of the recent topics in these $R$MnO$_3$ is the
presence of multiferroic properties.
For compounds with $R=$La, Nd, Sm, \ldots,
a spin A-type antiferromagnetic (A) phase is observed.
In this phase, spins order ferromagnetically in the $ab$ plane
and antiferromagnetically along $c$ axis, {\em i.e.},
the magnetic ordering vector is ${\bm q}=(0,0,2\pi)$ in the orthorhombic lattice with GdFeO$_3$-type distortions.
However, when one substitutes $R$ ions with smaller ones
$R=$ Tb, Dy, \ldots, 
an incommensurate spin spiral (ICS) phase emerges
accompanied by ferroelectricity.\cite{Kimura03,Kimura03a}
In this phase, magnetic propagation vectors are along the $b$ axis,
${\bm q}=(0,q_m,2\pi)$ where $0 < q_m < \pi$.
The origin of the 
ferroelectricity is 
the inverse Dzyaloshinsky-Moriya (DM) mechanism for the ICS structures.\cite{Katsura05,Sergienko06,Mostovoy06}
Here, a local electric polarization is generated through
the antisymmetric magnetoelectric (ME) coupling
$
 \bm P_{ij} \propto \bm e_{ij} \times (\bm S_i \times \bm S_j)$,
where $\bm S_i$ is the spin direction at $i$-th site and $\bm e_{ij}$ denotes the unit vector connecting $i$ and $j$ sites.
One of the experimental evidences is that the
reorientations of $\bm P$ from $\bm P \parallel a$ to $\bm P \parallel c$
by chemical substitutions or application of magnetic fields 
 are always accompanied by $ab$ to $bc$ cycloidal-plane flops.\cite{Yamasaki07,Yamasaki08}

Among various attempts to understand phase diagrams of $R$MnO$_3$,\cite{Kimura03a,Dong08}
finite temperature phase diagrams including $ab$- and $bc$-cycloid phases
have only been reproduced successfully by a model described as follows, so far: 
The model is a classical Heisenberg model defined on 
an three-dimensional orthorhombic lattice 
with nearest neighbor (n.n.) ferromagnetic exchanges $J_1$,
 and next nearest neighbor
(n.n.n.) antiferromagnetic exchanges $J_2$, as well as single ion anisotropies
and DM interactions.\cite{Mochizuki09,Mochizuki09a}
Models based on the $J_1$-$J_2$ Heisenberg model also explain spectra for magnons and electromagnons in these compounds.\cite{Miyahara08x,Aguilar09,Mochizuki10x}
Construction of a realistic and accessible 
microscopic model is quite appreciable
from the viewpoint of comprehension of mechanisms as well as
predictions and materials designs for novel phenomena.

\begin{figure}[h]
 \begin{center}
   \includegraphics[width=8cm,keepaspectratio,clip]{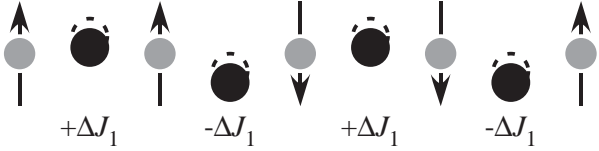}
  \end{center}
 \caption{
  Schematic view of the lattice distortion in E-type spin structure phase of $R$MnO$_3$. Gray circle with arrow shows Mn ions with spin $\uparrow\uparrow\downarrow\downarrow$ structure.
Displacement of oxygen ions (filled circle) gives rise to a uniform
electric polarization. Due to the Goodenough-Kamamori mechanism,
nearest neighbors exchanges are modulated as $\pm \Delta J_1$,
which gives a gain in total energy for E-type phase.
   }
 \label{fig:e-type}
\end{figure}

For smaller ions at $R$=Ho, Tm, Lu, \ldots, where $J_2/J_1$ is expected to be larger, 
spins exhibit an E-type antiferromagnetic (E) phase  with spin $\uparrow\uparrow\downarrow\downarrow$ collinear  structure
along the $b$-axis with the magnetic propagation vector ${\bm q}= (0,\pi,2\pi)$.
This phase also exhibits ferroelectricity $\bm P \parallel a$ and thus is multiferroic.
It is considered that the ferroelectricity 
is driven by the E phase orderings through
the symmetric ME coupling $|\bm P| \propto (\bm S_i \cdot \bm S_j)$ of the  magnetoelastic origin,\cite{Arima06,Sergienko07} as illustrated in Fig.~\ref{fig:e-type}.

The Heisenberg model with $J_1$-$J_2$ interactions alone, however,
 does not reproduce the E phase, as recently emphasized by 
 Kaplan.\cite{Kaplan09}
A possible way to stabilize the E phase is to introduce a uniaxial anisotropy which
enhances collinear behaviors. The authors have studied the Heisenberg model with  anisotropies and DM interactions, which successfully reproduces the A-ICS transition,
 in the large $J_2/J_1$ region.\cite{Mochizuki09a} 
Within the realistic range of parameters, however, we fail to observe the E phase.
Another candidate is the biquadratic interaction which also enhances collinearity.\cite{Kaplan09}
At this point, however, it is not clear weather such an interaction dominantly
acts to stabilize the E phase in $R$MnO$_3$.

In this paper, we introduce an alternative model to investigate the
spin structures of $R$MnO$_3$.
Namely, we study 
a classical Heisenberg model with bond alternations as a result of
the magnetoelastic couplings mentioned above.
We clarify that the model can reproduce the A-ICS-E phase transitions in $R$MnO$_3$, and discuss the nature of the ME phase diagram.
Since $R$MnO$_3$ is a rare system which exhibits
both symmetric and antisymmetric ME couplings
E and ICS phases, respectively,
it is also quite interesting
to study  the phase transition across these phases
in order to make further comprehension for the ME effects in these compounds.

\begin{figure}
 \begin{center}
   \includegraphics[width=8cm,keepaspectratio,clip]{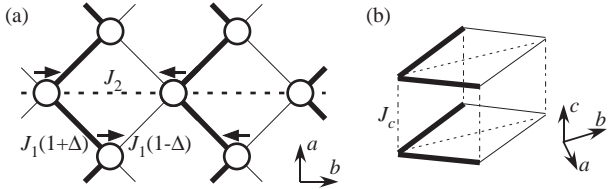}
 \end{center}
 \caption{
    Bond alternating $J_1$-$J_2$ model, depicted on (a)  the $ab$ plane
    and (b) the orthorhombic lattice.
     Thick and thin diagonal lines represent 
   nearest neighbor bonds with exchange coupling $(1+\Delta)J_1$ and $(1-\Delta)J_1$, respectively. Dotted lines represent bonds with $J_2$.
   Interlayer antiferromagnetic coupling $J_c$ is shown by dash-dotted lines.
   $a$, $b$ and $c$ represent orthorhombic axes. Thick arrows in (a) depict a spin configuration in the  E-type structure.
   }
 \label{fig:model}
\end{figure}

We study a classical Heisenberg model with spin exchange bonds depicted in Fig.~\ref{fig:model} as
\begin{align}
 {\cal H} &= - \sum_{<i,j>} J_{ij} \bm S_i \cdot \bm S_j 
  -J_2 {\sum_{<i,j>}}' \bm S_i \cdot \bm S_j   \nonumber\\
  & \qquad -J_c {\sum_{<i,j>}}'' \bm S_i \cdot \bm S_j.
\end{align}
Here, the first summation is taken over n.n. bonds on the $ab$ plane, and
$J_{ij}$ takes the alternating values along $b$-axis in the form
\begin{equation}
  J_{ij} = J_1 (1 \pm \Delta),
\end{equation}
as depicted in Fig.~\ref{fig:model}. 
Bond alternation parameter $\Delta$ is restricted within 
$0 \leqq \Delta \leqq 1$. 
The second  summation is
 for n.n.n. bonds along $b$-axis.
We take $J_1 >0$ (ferromagnetic) and $J_2 <0$ (antiferromagnetic).
The third term of the Hamiltonian
 is the  inter-plane antiferromagnetic couplings  along $c$-axis.
At $T=0$, $J_c$ merely create staggered stacking 
of $ab$-plane spin structure along $c$-axis, and thus
irrelevant for the phase diagrams with respect to $J_1$ and $J_2$.
For example, it is determined automatically that
a ferromagnetic alignment of spins in the $ab$ plane
stack antiferromagnetically  form the A-phase.
Therefore, we may focus on spin structures within the $ab$ plane.

In the absence of the bond alternation $\Delta=0$, the model at $T=0$ gives an ordinary
spiral state with $\theta_n = n\theta$, 
where $\theta_n$ 
describes the angle of the spin within the spiral plane.
Here, $n$ describes the position of the spin along the propagation vector $\parallel b$.
 The rotation angle $\theta$ is determined by $\cos\theta = 1/(2\gamma)$,
where 
\begin{equation}
   \gamma = - J_2 /J_1 ,
\end{equation}
and the $b$ component of the propagation vector is given by $q_m = 2\theta$.

\begin{figure}[tdp]
\includegraphics[scale=1.0]{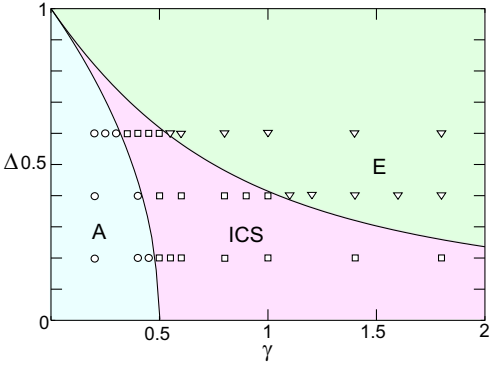}
\caption{(Color online) Phase diagram of the alternate $J_1$-$J_2$ model
at the ground state. A, E, and ICS denote A-type, E-type, and incommensurate spiral phases, respectively.  Solid lines are the phase boundaries determined by the analytical calculation. Circle, triangle, and square symbols denote the points at which the A, E, and ICS spin structures are respectively obtained in the Monte-Carlo calculation at $T/J_1$=0.2.}
 \label{fig:PhaseDiagram-Delta}
\label{Fig01}
\end{figure}

At $\Delta\ne0$, staggered modulation of $J_1$ is introduced.
Then we introduce a variational spin state with
uniform and staggered component for the rotation angle
\begin{equation}
  \theta_n = n\theta + (-1)^n \delta/2.
  \label{eq:theta_n}
\end{equation}  
Total exchange energy per site scaled by $J_1$  
 is described by $\varepsilon(\theta,\delta)$,
 which is calculated as
\begin{align}
  \varepsilon(\theta,\delta)
  &= -({1+\Delta})\cos(\theta-\delta) 
                               - ({1-\Delta}) \cos(\theta+\delta) \nonumber\\
                               & \qquad + \gamma \cos 2\theta - \gamma_c,
  \label{eq:energy1}
\end{align}
where $\gamma_c = -J_c/J_1$.
Minimization of  $\varepsilon(\theta,\delta)$ with respect to $\delta$ and $\theta$
through $\partial \varepsilon/\partial \theta =\partial \varepsilon/\partial \delta=0$ leads to
\begin{align}
  \tan\delta &= \Delta \tan \theta,  \label{eq:delta-theta}\\
  \cos\theta&=\sqrt{
       \frac{1-\Delta^2}{4\gamma^2} - \frac{\Delta^2}{1-\Delta^2}
          }.
    \label{eq:theta-gamma}
\end{align}
From Eq.~(\ref{eq:delta-theta})
we have $\delta=0$  in the limit
$\theta=0$ , which implies an A-phase, irrespective of $\Delta$. Similarly, at
$\delta=\pi/2$ we have $\theta=\pi/2$ which makes a collinear E phase.
 Critical value of $\Delta$ for the A-ICS boundary
is given from Eq.~(\ref{eq:theta-gamma}) at $\theta=0$ as
\begin{equation}
\Delta_c^{\rm (A)} = \sqrt{1-2\gamma} ,
\end{equation}
whereas at $\theta=\pi/2$ we have the ICS-E boundary
\begin{equation}
  \Delta_c^{\rm (E)} = \sqrt{\gamma^2+1}-\gamma.
\end{equation}
Since $\Delta_c^{\rm (E)}>0$, E phase is stabilized only in the presence of the bond alternation $\Delta$.
In Fig.~\ref{fig:PhaseDiagram-Delta} we show the phase diagram.


In order to justify the above analytical discussion, we perform the Monte-Carlo calculations at low temperatures. In the above, we have assumed the spin configuration of Eq.~(\ref{eq:theta_n}). On the other hand, the Monte-Carlo calculation does not restrict the spin configuration, and thus provides unbiased results. The calculation indeed confirms that only the three magnetic phases, i.e., A, ICS, and E phases are possible within our model so that the assumption of Eq.~(\ref{eq:theta_n}) is justified. As its consequence, the analytical results are precisely reproduced by the calculation.

In our numerical calculations,
we also add the  $\mathcal{H}_D=D\sum_{i}S_{z i}^2$, which makes magnetization along the $z$ axis hard. Because of this term, the spins in the ICS phase rotate in the $xy$ plane. 
This term makes the calculations stable by suppressing thermal fluctuations of the spiral plane without affecting the A-ICS-E transitions at $T=0$.
We take $J_c/J_1$=1 and $D/J_1$=0.2 in the calculation. We analyze this model using the Monte-Carlo technique for systems with 48$\times$48$\times$6 sites
with periodic boundaries.

The calculation successfully reproduces the analytically predicted phase diagram of Fig.~\ref{Fig01}. The circle, triangle and square symbols in Fig.~\ref{Fig01} denote the points at which the A, E, and ICS spin structures are respectively obtained in the Monte-Carlo calculation at $T/J_1$=0.2.

These magnetic structures are assigned from  the peak position of the
spin correlation functions at ${\bm q}=(0,q_m,2\pi)$.
We also measure the spin-helicity correlation function to identity the phases.
Here, the local spin-helicity vector is defined as
$\bm h_i=(\bm S_i \times \bm S_{i+\hat x}+\bm S_i \times \bm S_{i+\hat y})/2S^2$
where $\hat x$ and $\hat y$ point the n.n. sites along [110] 
(pseudo-cubic $x$) and [-110] (pseudo-cubic $y$) directions, respectively.
In  spiral structures, the rotating spins give rise to  a ferro-arrangement of the spin helicities,
which results in a peak of the spin-helicity correlation  at $\bm q$=0,
while in  collinear phases, 
the spin-helicity correlation  has no peak structure.

\begin{figure}[tdp]
\includegraphics[scale=1.0]{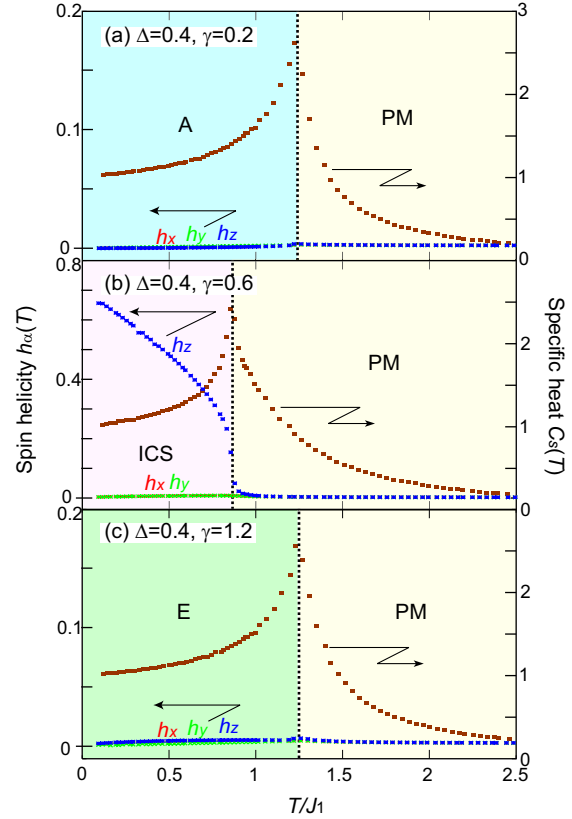}
\caption{(Color online) Calculated temperature profiles of specific heat $C_s(T)$ and spin-helicity vector $\bm h_s(T)$ for typical points in the phase diagram where the ground states are estimated to be (a) A-type, (b) ICS, and (c) E-type. 
PM denotes the paramagnetic phase. }
\label{Fig03}
\end{figure}

Let us now show some details of the calculation.
In Fig.~\ref{Fig03}, we show the temperature dependence 
of specific heat $C_s(T) =(1/N) ( {\partial \langle \mathcal{H}\rangle} )/({\partial (k_{\rm B}T)})$,
as well as the spin-helicity vector $\bm h_s(T)$ where
$h_\alpha(T)$ ($\alpha$=$x$, $y$ and $z$) denotes the $\alpha$ component of the spin-helicity vector, at (a) ($\Delta$, $\gamma$)=(0.4, 0.2), (b) ($\Delta$, $\gamma$)=(0.4, 0.6), and (c) ($\Delta$, $\gamma$)=(0.4, 1.2) where the ground states are predicted to be (a) A-type, (b) ICS, and (c) E-type, respectively ----- see also the phase diagram in Fig.~\ref{Fig01}. 

In all cases, we can see a single phase transition from paramagnetic (PM) phase at high temperatures to each ordered phase at low temperatures, at which $C_s(T)$ exhibits a sharp peak. 
We also confirm that the choice of $T/J_1=0.2$ gives a sufficiently low-temperature state.

 In the present calculation, the spins in the ICS phase rotate in the $xy$ plane since we incorporate the hard-axis type spin anisotropy along the $z$ axis in the Hamiltonian. Under this circumstance, $h_z(T)$ has a large value, while $h_x(T)$ and $h_y(T)$ are almost zero. We can indeed see that $h_z(T)$ starts increasing at the transition to the ICS phase in Fig.~\ref{Fig03}(b). On the other hand, as shown in Figs.~\ref{Fig03}(a) and ~\ref{Fig03}(c), 
 $z$ component of the helicity  is as small as other components in the collinear spin phases.
\begin{figure}[tdp]
\includegraphics[scale=1.0]{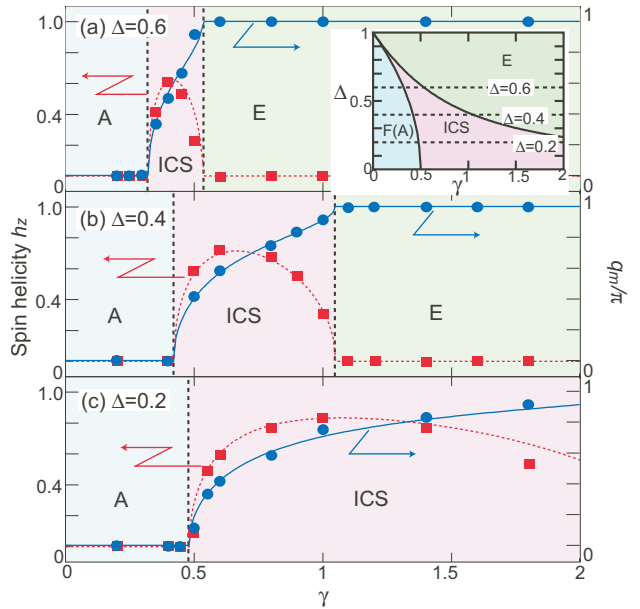}
\caption{(Color online) Calculated $\gamma$ dependence of the momentum $q_m/\pi$ at which the spin correlation function has a peak at $T/J_1$=0.2, and $\gamma$ dependence of the spin-helicity $z$ component $h_z$ at $T$=0 for (a) $\Delta$=0.6, (b) $\Delta$=0.4, and (c) $\Delta$=0.2. The data of $h_z$ at $T$=0 are obtained by extrapolating the calculated temperature profile of $h_z(T)$. Solid and dashed lines are  $q_m/\pi$ and $h_z$
obtained from the analytical approach.}
\label{Fig04}
\end{figure}

In Fig.~\ref{Fig04}, we show the calculated $\gamma$ dependence of the momentum $q_m/\pi$ as well as the spin-helicity  $h_z$ 
for (a) $\Delta$=0.6, (b) $\Delta$=0.4, and (c) $\Delta$=0.2.
The data for the spin-helicity are obtained by extrapolating the calculated temperature profile of $h_z(T)$ to $T/J_1\to0$. 
Those data as functions of $\gamma$ consistently
exhibit A-ICS-E phase transitions, at least qualitatively.

Furthermore, let us compare these data with the variational
spin structure given in Eq.~(\ref{eq:theta_n}).
The spin structure gives $q_m=2\theta$ and 
$h_z=\sin\theta\cos\delta$,
where $\theta$ and $\delta$ are determined by Eqs.~(\ref{eq:delta-theta}) and (\ref{eq:theta-gamma}). As plotted in 
Fig.~\ref{Fig04}, the Monte-Carlo data for $q_m$ and $h_z$ agree well with the analytical results given by solid and dashed lines, respectively.


So far we have confirmed that the spin structure given in Eq.~(\ref{eq:theta_n})
as well as  the A-ICS-E phase transition derived from that is quite valid.
Then we proceed to consider the case that the bond alternation is driven by 
 lattice distortions to minimize the total energy.
We assume that a  
lattice distortion $r$ creates the bond alternation $\Delta = g r$
with $g$ being the coupling constant.
Total energy is given by a sum of  the spin exchange energy $\varepsilon_{\rm s}(\Delta,\gamma)$  and 
an elastic energy $ \propto r^2$, or equivalently
\begin{equation}
 \varepsilon_{\rm tot}
   = \varepsilon_{\rm s}(\Delta,\gamma) + \frac{\Delta^2}{2\lambda^2} .
\end{equation}
Here, $\lambda$ is the dimensionless spin-lattice coupling constant.
Hereafter, we refer to the model as Peierls $J_1$-$J_2$ model.

The spin exchange energy is derived by applying Eqs.~(\ref{eq:delta-theta}) and (\ref{eq:theta-gamma}) to Eq.~(\ref{eq:energy1}).
In the A phase at $\Delta <\Delta_c^{\rm(A)}$, the minimized energy is derived from $\theta=\delta=0$ as
\begin{equation}
  \varepsilon_{\rm s}(\Delta,\gamma)= -2 +{\gamma} -\gamma_c,
\end{equation}  
while in the E phase at $\Delta > \Delta_c^{\rm(E)}$,   we have
$\theta=\delta=\pi/2$ so that
\begin{equation}
   \varepsilon_{\rm s}(\Delta,\gamma)= - 2\Delta -{\gamma} -\gamma_c.
\end{equation}
In the ICS phase at $\Delta_c^{\rm (A)}<\Delta<\Delta_c^{\rm (E)}$, we have
\begin{equation}
 \varepsilon_{\rm s}(\Delta,\gamma) =
  - \frac{1-\Delta^2}{2\gamma} + \gamma\frac{1+\Delta^2}{1-\Delta^2} -\gamma_c .
\end{equation}

\begin{figure}
 \begin{center}
   \includegraphics[width=6.5cm,keepaspectratio,clip]{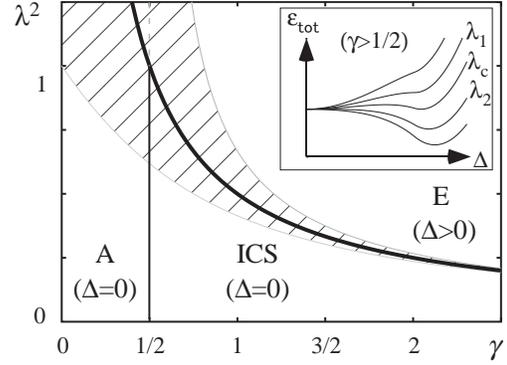}
 \end{center}
 \caption{
   Phase diagram of the Peierls $J_1$-$J_2$ model. 
  The solid curve shows the first-order phase boundary between E phase  and ICS (F) phase  at $\gamma>1/2$ ($\gamma<1/2$).
   The vertical solid line at $\gamma=1/2$ shows the second-order phase boundary between ICS and F phases.
    Hatched area shows the coexistence region.
     Inset: Total energy $\varepsilon_{\rm tot}$ as a function of $\Delta$ plotted in arbitrary units, for the case of $\gamma >1/2$. Dimensionless coupling constant $\lambda$ is
     increased from top to bottom, as $\lambda < \lambda_1$, $\lambda=\lambda_1$, $\lambda=\lambda_c$, $\lambda=\lambda_2$ and $\lambda > \lambda_2$, respectively.
   }
 \label{fig:PhaseDiagram-lambda}
\end{figure}

Bond alternation $\Delta$ is determined by minimizing $\varepsilon_{\rm tot}$
 with respect to  $\Delta$ within the range
$0 \leqq \Delta \leqq 1$.
The result in the region $\gamma > 1/2$ can be summarized as follows.
In Fig.~\ref{fig:PhaseDiagram-lambda}, we schematically depict $\varepsilon_{\rm tot}$ as a function of $\Delta$ 
for various $\lambda$.
We observe a transition from ICS phase to E phase
at a critical coupling $\lambda_{\rm c}$ given by
\begin{equation}
  \lambda_{\rm c}{}^2 = 1/({2\gamma}).
  \label{eq:lambda_c}
\end{equation}
Here, $\varepsilon_{\rm tot}$ have degenerate energy minimums
at $\Delta=0$ (ICS region) and at $\Delta > \Delta_c^{\rm(E)}$ (E region).
As $\lambda$ is changed across $\lambda_{\rm c}$, 
we have a jump in the value of $\Delta$ for the energy minimum and thus the transition is first ordered.
Furthermore, there exists a characteristic coupling
\begin{equation}
  \lambda_{1}{}^2 = \sqrt{\gamma^2+1}-\gamma,
   \label{eq:lambda_1}
\end{equation}
where  E phase has a local minimum
in its energy with respect to $\Delta$ at $\lambda > \lambda_1$.
This implies  an existence
of a metastable E state at $\lambda > \lambda_1$.
Similarly, we also have 
\begin{equation}
  \lambda_{2}{}^2 = 2{\gamma}/({4\gamma^2-1}),
 \label{eq:lambda_2}
\end{equation}
where an ICS state at $\Delta=0$ is metastable at $\lambda < \lambda_2$.
Therefore, in the case $\gamma>1/2$, we may have coexistence of the ICS and the E phases at $\lambda_1 < \lambda < \lambda_2$.

Similarly, in the case $\gamma \leqq 1/2$, 2nd order transition between the A and the ICS phases 
is observed at $\lambda_c$, while the metastable E state exist at $\lambda > \lambda_1$. The A states is always metastable at $\lambda>\lambda_c$ in this case.
The forms for $\lambda_c$ and $\lambda_1$ coincide with those
for the previous case given in Eqs.~(\ref{eq:lambda_c}) and (\ref{eq:lambda_1}), respectively.  
The phase diagram is summarized in Fig.~\ref{fig:PhaseDiagram-lambda}.

 Let us now compare these results with the magneto-electric properties in $R$MnO$_3$.
In these compounds,  commensurate E phase is 
always accompanied with ferroelectricity,
which is consistent with our result that E phase is stabilized 
only at $\Delta\ne0$.
Although it has previously been discussed that ferroelectricity is
triggered by magnetic E phase through lattice distortions,
our result suggests that the magnetic E phase and ferroelectricity
{\em cooperatively} emerge to stabilize themselves.
Therefore,  Peierls-type spin-lattice couplings are essentially important to
understand the electric as well as the magnetic phase diagram of $R$MnO$_3$.
Recent report shows that, in $R$MnO$_3$ there exists 
a region of possible coexistence for ICS and E phases.\cite{Ishiwata09x}
The Peierls model discussed here indeed shows such a coexistence at
around the 1st-order transition points between ICS and E phases.
Experimental data in further details should give crucial tests
for our present results. 
Theoretical approaches based on more realistic models
including anisotropies, DM and biquadratic interactions 
should also
be important to understand the whole phase diagrams
of $R$MnO$_3$ in detail.

The author would like to thank S. Miyahara
for stimulating discussions, as well as
S. Ishiwata, Y. Taguchi and Y. Tokura for suggestions from experimental points of view.
This work was partially supported by Grant-in-Aid for Scientific Research
as well as High Tech Research Center Project
 from the MEXT, Japan.


\begin{thebibliography}{99}

\bibitem{Tokura00}
For a review, {\em ``Colossal Magnetoresistive Oxides''}, ed. by
Y. Tokura (Gordon \& Breach Science Publisher, 2000).
\bibitem{Kimura03}
T. Kimura, T. Goto, H. Shintani, K. Ishizaka, T. Arima and Y. Tokura: Nature {\bf 426} (2003) 55.

\bibitem{Kimura03a}
T. Kimura, S. Ishihara, H. Shintani, T. Arima, K. T. Takahashi, K. Ishizaka, and Y. Tokura:
Phys. Rev. B {\bf 68} (2003) 060403.

\bibitem{Katsura05}
H. Katsura, N. Nagaosa, and A. V. Balatsky: Phys. Rev. Lett. {\bf 95} (2005) 057205.

\bibitem{Sergienko06} I. A. Sergienko and E. Dagotto: Phys. Rev. B 73 (2006) 094434.
\bibitem{Mostovoy06} M. Mostovoy: Phys. Rev. Lett. 96 (2006) 067601.


\bibitem{Yamasaki07}
Y. Yamasaki, H. Sagayama, T. Goto, M. Matsuura, K.
Hirota, T. Arima, and Y. Tokura: Phys. Rev. Lett. {\bf98} (2007)
147204.

\bibitem{Yamasaki08}
Y. Yamasaki, H. Sagayama, N. Abe, T. Arima, K. Sasai, M. Matsuura, K. Hirota, D. Okuyama,
Y. Noda, and Y. Tokura: Phys. Rev. Lett. {\bf101} (2008) 097204.


\bibitem{Dong08}
S. Dong, R. Yu, S. Yunoki, J.-M. Liu, and E. Dagotto:
Phys. Rev. B {\bf78} (2008) 155121.

\bibitem{Mochizuki09}
M. Mochizuki and N. Furukawa:
J. Phys. Soc. Jpn. {\bf 78} (2009) 053704.

\bibitem{Mochizuki09a}
M. Mochizuki and N. Furukawa:
Phys. Rev. B {\bf 80} (2009) 134416.

\bibitem{Miyahara08x}
S. Miyahara and N. Furukawa:
preprint, arXiv:0811.4082.

\bibitem{Aguilar09}
R. Vald{\'e}s Aguilar, M. Mostovoy, A. B. Sushkov, C. L. Zhang, Y. J. Choi, S-W. Cheong, and H. D. Drew:
Phys. Rev. Lett. {\bf102} (2009) 047203.


\bibitem{Mochizuki10x}
M. Mochizuki, N. Furukawa and N. Nagaosa: arXiv:1001.3905.

\bibitem{Arima06}
T. Arima, A. Tokunaga, T. Goto, H. Kimura, Y. Noda, and Y. Tokura:
Phys. Rev. Lett. \textbf{96} (2006) 097202.

\bibitem{Sergienko07}
Ivan A. Sergienko, Cengiz Sen and Elbio Dagotto:
Phys. Rev. Lett. \textbf{97} (2006) 227204.


\bibitem{Kaplan09}
T.A. Kaplan:
Phys. Rev. B{\bf 80} (2009) 012407.

\bibitem{Ishiwata09x}
S. Ishiwata, Y. Kaneko, Y. Tokunaga, Y. Taguchi, T. Arima and Y. Tokura: 
arXiv:0911.4190.
\end{thebibliography}
\end{document}